\documentclass[aip,reprint,longbibliography]{revtex4-1}
\usepackage{graphicx}
\usepackage{color}
\usepackage{amsmath,amssymb,tikz}
\usepackage{placeins}
\usepackage{pifont}
\usepackage{amsfonts}
\usepackage{amsmath}
\usepackage{comment}

\draft 

\begin{document}

\title{Diffusion and viscosity of  non-entangled polyelectrolytes.} 


\author{Carlos G. Lopez}
\email[]{lopez@pc.rwth-aachen.de}
\affiliation{Institute of Physical Chemistry, RWTH Aachen University, Landoltweg 2, 52056 Aachen, Germany}
\author{J\"{u}rgen Linders}
\author{Christian Mayer}
\email[]{christian.mayer@uni-due.de}
\affiliation{Physical Chemistry and Center for Nanointegration Duisburg-Essen (CeNIDE), University of Duisburg-Essen, 45117 Essen, Germany}
\author{Walter Richtering}
\affiliation{Institute of Physical Chemistry, RWTH Aachen University, Landoltweg 2, 52056 Aachen, Germany}





\begin{abstract}
We report chain self-diffusion and viscosity data for sodium polystyrene sulfonate (NaPSS) in semidilute salt-free aqueous solutions measured by pulsed field gradient NMR and rotational rheometry respectively. The observed concentration dependence of $\eta$ and $D$ are consistent with the Rouse-Zimm scaling model  with a concentration dependent monomeric friction coefficient. The concentration dependence of the monomeric friction coefficient exceeds that expected from free-volume models of diffusion, and its origin remains unclear. Correlation blobs and dilute chains with equivalent end-to-end distances  exhibit nearly equal friction coefficients,  in agreement with scaling. The viscosity and diffusion data are combined using the Rouse model to calculate the chain dimensions of NaPSS in salt-free solution, these  agree quantitatively with SANS measurements.
\end{abstract}

\maketitle 

\section{Introduction}\vspace{-0.4cm}
Polyelectrolytes are essential components of living systems, where they form interpenetrating networks that impart  mechanical integrity to biological tissue. Understanding the structure and dynamics  of polyelectrolytes networks is an important step towards unravelling  complex transport phenomena in biological environments, which underpin many physiological processes such as muscle contraction, nerve signalling or joint lubrication.\cite{horkay2000osmotic,schwörer2018drastic,ballauff2018more} 

The strong, long ranged electrostatic interactions between like-charged  chains means that polyelectrolyte solutions and gels are highly correlated systems\cite{yethiraj2009liquid,chang2015osmotic,muthukumar201750th,RubinsteinReview,dedic2019polyelectrolytes,chremos2018polyelectrolyte,chremoscomparison,morishima2018small,Jia2019Effect,Muthu2019,chen2020influence}, for which the success of mean field theories (e.g. the random phase approximation) is limited, particularly in low ionic strength solvents.\cite{barrat1996theory,RubinsteinReview} Scaling models, following the pioneering work of de Gennes and co-workers\cite{deGennes,pfeuty1978conformation,noda1996viscoelastic,Dobrynin,RubinsteinRouseSimulation,DobryninSaltSimulation}, provide a relatively simple and useful framework to understand polyelectrolyte behaviour, and are particularly successful in predicting the conformational properties of flexible polyelectrolytes in solution.\cite{PSSI} The scaling treatment of polyelectrolyte dynamics on the other hand displays poorer agreement with experimental results.\cite{ColbyPSS,MaleateSANS,PSSI,PSSII,PSSIII,CMCLetter}

In salt-free or low ionic strength media, polyelectrolytes adopt highly extended conformations\cite{StevensSimulation,PSSI,CMCII,CMCLetter,mintis2019effect,PSSIII,mintis2019effect2,soysa2015size,mantha2015conformational,xu2016single,gupta2019structure}, meaning that their solutions are  above the overlap concentration ($c^*$) for most practical applications. While the overlap concentration of polyelectrolytes is  much lower than that of non-ionic polymers, especially for high degrees of polymerisation ($N$), the entanglement concentration ($c_e$) is only weakly dependent on charge fraction\cite{Dou2006,PSSIII}. As the result, there exists a wide  range in the $N-c$ phase space where polyelectrolytes are in the semi-dilute non-entangled regime ($c^* < c < c_e$).\cite{PSSI,PSSIII,CMCLetter} For example,  sodium polystyrene sulfonate (NaPSS) with $N \simeq 2000$ displays  $c_e/c^*\simeq 10^3$ in salt-free water. By comparison, $c_e/c^*\simeq 10$ for polystyrene with equivalent degree of polymerisation in good solvent.\cite{PSSIII}

Single chain diffusion measurements provide an important test to the microscopic description of polymer dynamics put forward  by various theories\cite{vagias2013complex,seiffert2008diffusion}. In the context of polyelectrolytes,  chain diffusion has been studied primarily in the excess salt, zero polymer concentration limit by  dynamic light scattering\cite{serhatli2002coil,yashiro2002excluded,sedlak2017poly}, as well as by NMR, fluorescence methods and centrifugation in the infinite polymer dilution limit, both in salt-free and excess added salt\cite{PSSI,bohme2007effective,scheler2009nmr,bohme2007hydrodynamic,cao2015dynamics,wandrey2016hydrodynamic,wang2015investigation}. Measurements above $c^*$ have received far less attention\cite{filippov2013collective,oostwal1993novel,oostwal1993chain}, and most experimental data come from two studies by Oostwal and co-workers\cite{oostwal1993novel,oostwal1993chain}, who investigated the molar mass, polymer concentration and added salt concentration variation of the diffusion coefficient of NaPSS in D$_2$O. Studies on  solvent, counter-ion and co-solute diffusion in polyelectrolyte and ionomer solutions and gels have also been reported. \cite{schipper1996polyelectrolyte,schipper1998influence,schneider2006solvent,senanayake2019diffusion,guo2019dynamics,guo2019structure,sozanski2016motion,lafemina2016diffusion,bai2014nmr}

Rheological data, particularly the zero shear-rate viscosity of solutions, provide a complementary test to the macroscopic dynamics of polyelectrolytes \cite{noda1996viscoelastic,ColbyPSS,PSSI,PSSII,Dou2006,Dou2008,MaleateSANS}. Due to the relative ease with which such measurements can be carried out, a vast literature on the viscosity of polyelectrolytes  exists, covering a wide range of polyelectrolyte types\cite{zheng2018counterion,dakhil2019infinite,fouissac1993shear}, molar masses\cite{ColbyPSS,fouissac1993shear,Izzo2014the,bravo2016conformation}, charge densities\cite{chen2019viscosity,chen2020the,konop1999polyelectrolyte}, solvent quality\cite{Waigh2001}, dielectric constant\cite{Dou2008,jimenez2018extensional,rozanska2019capillary,de2020concentration} and added salts\cite{MaleateDynamic,turkoz2018salt,Bercea2018Intrinsic,bercea2019associative,Pavlov2018Spectrum,jimenez2020capillary,CMCLetter}.

The viscosity of  polyelectrolyte solutions was first studied by Fuoss  and co-workers  \cite{fuoss1948viscosity,fuoss1949viscosity}, who observed a linear dependence of the reduced viscosity ($\eta_{red}$) and the on the square root of the polymer concentration for poly-4-vinylpyridine derivatives in water-ethanol mixtures. Similar behaviour was  observed for many other systems and eventually became known as the Fuoss law\cite{ColbyReview,PSSII}. A theoretical explanation for this unusual dependence was first put forward by de Gennes' and co-workers\cite{deGennes}, and later incorporated into the theories of Yamaguchi et al\cite{yamaguchi1992viscoelastic} and Dobrynin et al\cite{Dobrynin} among others. Boris and Colby\cite{ColbyPSS} identified significant deviations from the Fuoss law for NaPSS in DI water, and after a careful review of literature data concluded that much (but not all) of the observed behaviour could be attributed to artefacts related to shear thinning and/or salt contamination.\cite{ColbyPSS} Deviations to stronger power-laws have also been reported for other systems\cite{CMCII,CMCLetter,JSB,del2017relaxation,kujawa2006effect}. Recently, an extensive analysis of literature data for  the viscosity of non-entangled NaPSS\cite{PSSII} in salt-free water showed that the power-law exponent of the specific viscosity with concentration increases with increasing polymer concentration, and agrees with the Fuoss law only for $c \lesssim 0.05$ M. The  origin of the concentration dependence of the viscosity-concentration exponent could however not be established as viscosity and diffusion data were in apparent conflict.\cite{PSSI,PSSII,oostwal1993novel,oostwal1993chain}

While it is common to treat solvent dynamics as independent of polymer concentration, it is known that the addition of polymers, salts, co-solvents or nanoparticles alters the cohesive energy between solvent molecules\cite{gong2001self,bai2014nmr,DouglasProtein,DouglasCG,DouglasSalts,DouglasHF,DouglasNP}. The structuring of water by solutes with ionic groups in particular has received renewed attention over the last few years, as recent results indicate that nuclear quantum effects   \cite{belloni2018screened,Jungwirth2018Ion,WaterOrientation1}. This leads to changes the thermodynamic and transport properties of solvents, a feature that is not considered by continuum theories such as Rouse-Zimm models. Experimental findings show that addition of polymers to a solvent leads to a slow-down of solvent dynamics, which can be captured based on obstruction or free-volume models.\cite{MM,Fujita1991,colby1991effects} Other effects not included in the scaling theories are the influence of counterion solvation and polyelectrolyte clustering\cite{tarokh2019atomistic,SedlakSlowMode1,sedlak1999can,sedlak1992concentration,chremos2018polyelectrolyte,chremos2018competitive,narayanan2018specific}. The impact of these on the dynamics of polyelectrolytes is not well understood at present.

The purpose of this work is to present new results for the viscosity and diffusion coefficient of NaPSS in salt-free solution as a function of polymer concentration and molar mass, and to compare these with scaling predictions. The paper is organised as follows: we first outline the scaling theory of non-entangled salt-free polyelectrolytes and  review the available experimental data. We then present diffusion and viscosity results for NaPSS and show that scaling can account for most experimental observations  if a concentration dependent monomeric friction coefficient is assumed. Possible explanations for the strong concentration dependence of the monomeric friction factor are considered. We discuss significance of the product of the viscosity increment and the diffusion coefficient, and  show that it can be used to obtain quantitative estimates of the  dimensions of non-entangled chains. Some observations, particularly a discrepancy between the theoretical and experimental $N-$ dependence of the specific viscosity\cite{PSSI} remain unexplained.\vspace{-0.5cm}
\section{Literature Review}\vspace{-0.3cm}

\subsection{Polyelectrolyte Scaling}\label{sec:Scaling}\vspace{-0.4cm}

Polyelectrolytes in dilute solution adopt a rod-like conformation with an end-to-end distance of $R \simeq b'N$, where $b'$ is the effective monomer size and $N$ the degree of polymerisation.\cite{Dobrynin,ColbyPSS,PSSIII,NoteBlob,CMCLetter} The overlap concentration scales as $c^* \simeq N/R^3 \simeq b'^3N^{-2}$. For $c > c^*$ polyelectrolytes interpenetrate, forming a mesh with size $\xi \simeq b'^3c^{-1/2}$, also known as the correlation length. Their conformation is a random-walk of $g = (c/c^*)^{1/2}$ correlation blobs, with an end-to-end distance of $R \simeq \xi (L/\xi)^{1/2} \simeq b'^{1/4}N^{1/2}c^{-1/4}$.\cite{PSSI}

The scaling theory of de Gennes and co-workers' treats semidilute polyelectrolyte chains as Rouse chains made up of correlation blobs.\cite{deGennes} In salt-free solution, each correlation blob assumes a rod-like conformation with diameter $d_C$ and length $\xi$. The scaling theory approximates the friction coefficient as:\vspace{-0.2cm}
\begin{equation}
\zeta_\xi =(F \beta) \xi
\label{eq:Fxi}
\end{equation}

\noindent where $F$ is a shape factor ($F = 6\pi$ for spheres) and $\beta$ is a coefficient related to the local friction experienced by polymer chains. Scaling assumes $F = 1$ and $\beta = \eta_s$, where $\eta_s$ is the viscosity of the solvent.\cite{Dobrynin} 

The Rouse model expects to total friction coefficient to be that of a single correlation blob multiplied by the number of blobs in a chain:\vspace{-0.2cm}
\begin{equation}
\zeta_R \simeq \frac{Nb'}{\xi}\zeta_\xi = (F\beta) b'N
\label{eq:Fchain}
\end{equation}
\noindent which Dobrynin et al's model expects to be concentration independent.\cite{Dobrynin,RubinsteinReview,ColbyReview}

The Rouse diffusion coefficient of a chain is obtained via the Einstein-Smoluchowski relation:\vspace{-0.2cm}
\begin{equation}
D_R = k_BT/\zeta_R \simeq (F\beta)^{-1}\frac{k_BT}{Nb'}
\label{eq:DRouse}
\end{equation}

\noindent where $k_B$ and $T$ are the Boltzmann constant and the absolute temperature respectively. 

The chain's relaxation time  is $\tau_R \simeq R^2/D$ so that:
 \begin{equation}
\tau_R \simeq
\left\{
\begin {array}{cccc}
 (F\beta) \frac{b'^{3/2}N^2c^{-1/2}}{k_BT} \  \ \ \ \ \ \ \ \ \ \ \text{scaling} \\ 
\noalign{\medskip}
  \frac{1}{29.6}(F\beta) \frac{b'^{3/2}N^2c^{-1/2}}{k_BT}  \  \ \ \  \ \ \ \text{Rouse}
\end {array}
\right.
\label{eq:tauRouse}
\end{equation}

\noindent where the factor of  $3\pi^2 \simeq 29.6$ in Eq. \ref{eq:tauRouse}b was calculated by Rouse, and is not usually included in  scaling theories.

In the absence of entanglements, the terminal modulus of a single chain is $\simeq k_BT$ and that of a solution is: $G \simeq \frac{k_BTc}{N}$.\cite{Dobrynin,PSSI} The polymer contribution to the viscosity is estimated as the product of $\tau_R$ and $G$:
 \begin{equation}
\eta - \eta_s  \simeq
\left\{
\begin {array}{cccc}
 (F\beta) b'^{3/2}Nc^{1/2} \  \ \ \ \ \ \ \ \ \ \ \text{scaling} \\ 
\noalign{\medskip}
  \frac{1}{36}(F\beta) b'^{3/2}Nc^{1/2}  \  \ \ \  \ \ \ \text{Rouse}
\end {array}
\right.
\label{eq:etaRouse}
\end{equation}

Equations \ref{eq:Fxi}-\ref{eq:etaRouse} with $F = 1$ and $\beta = \eta_s$ correspond to Dobrynin et al's scaling model. In section \ref{sec:discussion} we evaluate the value of these parameters as a function of polymer concentration and molar mass.\vspace{-0.5cm}
\subsection{Experimental results on polyelectrolytes}\vspace{-0.5cm}

We now turn our attention to the experimental results from the literature, specifically for NaPSS in water, the most extensively studied polyelectrolyte system. The conformational properties of NaPSS in salt-free solution are in nearly quantitative agreement with the scaling predictions\cite{PSSI}, but larger deviations from theory are observed for dynamic quantities.\cite{ColbyPSS,PSSI,PSSII,PSSIII}  Of particular interest for this study is the fact that the specific viscosity of NaPSS shows a strong increase with polymer concentration beyond $c \simeq 1$ M, displaying exponential behaviour that is not anticipated by Dobrynin et al's model.\cite{PSSII}

We define a function $\psi$ that accounts for deviations from the scaling prediction as follows:
\begin{equation}
\psi_X(c,N) \equiv \frac{X_E(c,N)}{X_T(c,N)}
\label{eq:psi}
\end{equation}

\noindent where $X$ refers to the solution property (e.g. $X = \eta_{sp}$, $D$), and the subscripts $E$ and $T$ refer to the experimental result and the theoretical value respectively. 

For $X = \eta_{sp}$, two recent studies\cite{PSSI,PSSII} of semidilute, non-entangled NaPSS solutions  showed that $\psi$ could be written as $\psi_\eta(c,N) = A_0\psi_\eta(N)\psi_\eta(c)$, where 
$$
\psi_\eta(N) \simeq N^{0.25 \pm 0.05}
$$ \vspace{-0.25cm}
\noindent and \vspace{-0.1cm}
$$
\psi_\eta(c) \simeq (15/16)e^{1.4c} + e^{1.3c^2}/16
$$
\noindent where $c$ in units of moles of repeating units per dm$^3$ and $A_0$ is a $c-$ and $N-$ independent constant of order unity.\cite{PSSI,PSSII}

The function $\psi_{D}(c,N)$ displayed a more complex dependence, with the $c$ and $N$ parts not being separable. At low polymer concentrations $\psi_D(c,N) \simeq 1$ is observed. The divergence between $\psi_{D}(c)$ and $\psi_{\eta}(c)$ is somewhat surprising, as the simplest explanation for a $N-$independent slow-down of non-entangled polymer dynamics at high concentrations is an increase in the local friction experienced by polymer chains, which would affect both functions in precisely an inverse manner (i.e. $\psi_D(c) = [\psi_\eta(c)]^{-1}$).

In contrast to $\psi_\eta(c,N)$, which was determined from data originating from nearly 20 different literature sources\cite{PSSI,PSSII}, the evaluation of $\psi_{D}(c,N)$ was carried out using the results of  only two  studies. In the current paper, we present new data for the diffusion coefficient of NaPSS in salt-free solution in order to compute $\psi_{D}(c,N)$ over a wider range of $c$ and $N$. We show that the $\psi_\eta(c)$ and $\psi_D(c)$ functions are consistent with the scaling theory coupled with a $c-$dependent $\beta$, but the origin of this dependence remains unclear.

\vspace{-0.1cm}
 \section{Materials and Methods}\vspace{-0.3cm}
\noindent{\bf Materials:} Sodium polystyrene sulfonates with weight-averaged molar masses of $M_w =$ 9.7, 14.9, 20.7, 29.2, 63.9, 148, 151 and 259  kg/mol were purchased from Polymer Standard Services (Mainz, Germany). DI water was obtained from a Milli-Q source. D$_2$O was purchased from Sigma-Aldrich (conductivity 2 $\mu$S/cm. Samples were prepared gravimetrically and stored in plastic vials.

\begin{table}[h]
\begin{tabular}{l|llll}
$M_w$ (g/mol)  & $c^*_\eta$ (M)$^a$& $c^*$ (M)$^b$ & $pd^c$ \\ \hline
$9.7 \times 10^3$ &   & 5.1 $\times 10^{-1}$ & 	1.04 \\
$1.49 \times 10^4$ &  & 2.2 $\times 10^{-1}$ & 1.04\\
$2.07 \times 10^4$ &  &1.1 $\times 10^{-1}$ & 1.05\\
$2.91 \times 10^4$ & 5.9 $\times 10^{-2}$ & 5.7 $\times 10^{-2}$ & 1.06\\
$6.39 \times 10^4$ &  1.2 $\times 10^{-2}$ & 1.2 $\times 10^{-2}$ & 1.04\\
$1.48 \times 10^4$ & 2.0 $\times 10^{-3}$ & 2.2 $\times 10^{-3}$ & 1.04 \\
$2.61 \times 10^4$ & 4.9 $\times 10^{-5}$ & 7.0 $10^{-5}$ & 1.02 \\
\end{tabular}
\caption{Molar masses of sodium polystyrene sulfonate used. $^a$ $c_\eta^*$ is determined from the $\eta_{sp}(c^*) =0.67$ criterion, as detailed in ref. \citenum{PSSIII}. $^b$ Estimated as $c^* = 1240 N^{-1.8}$, see ref. \citenum{PSSIII}. Data for the four highest molar masses are from ref. \citenum{PSSII}. $^c$ Value for  polystyrene before sulfonation. } 
\label{tab:Table1}
\end{table}
\noindent{\bf Rheology:} A Kinexus-Pro rheometer with a cone-and-plate geometry (40 mm diameter, 1$^\circ$ angle) or a plate-plate geometry (20 mm, gap 0.4 mm gap) was employed for the steady and oscillatory shear experiments. 

\noindent{\bf Pulsed-Field Gradient Nuclear Magnetic Resonance (PFG-NMR):} $^1$H NMR diffusion experiments were run on a 500 MHz Bruker Avance NEO II spectrometer with a Bruker DIFFBBI probe head. All measurements were performed at 298.15 K. 
The polymers were dissolved in D$_2$O to suppress the water signal and to  obtain a lock signal in the NMR experiment. 5 mm NMR tubes were filled with 500 $\mu$L of polymer solution. For all measurements, the stimulated echo pulse sequence combined with two gradient pulses was used. Thirty-two scans were accumulated for each gradient  setting. The time delay between two gradient pulses  $\Delta$ was set to 25 ms. The gradient pulses were adjusted to strengths $G$ between 5 and 500 G/cm with a duration $\delta =$ 1.0 ms. All measurements (the full set of gradient strengths under the variation from 5 to 500 G/cm) were repeated four times. 
\vspace{-0.5cm}
\section{Results and Data Analysis}\vspace{-0.4cm}
\begin{figure}[b!]
\includegraphics[width=3in]{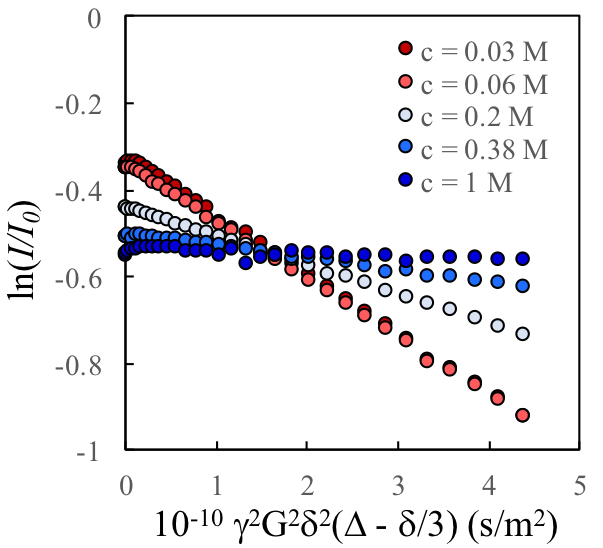}\vspace{-0.35cm}
\label{fig:Results2}
\caption{Stejskal-Tanner plot obtained from the PFG-NMR measurements for NaPSS with $M_w = 64$ kg/mol at varying concentration. The slope of the straight line gives the negative value of the diffusion coefficient of the investigated molecule.}
\end{figure}

\begin{figure*}[ht]
\includegraphics[width=6.3in]{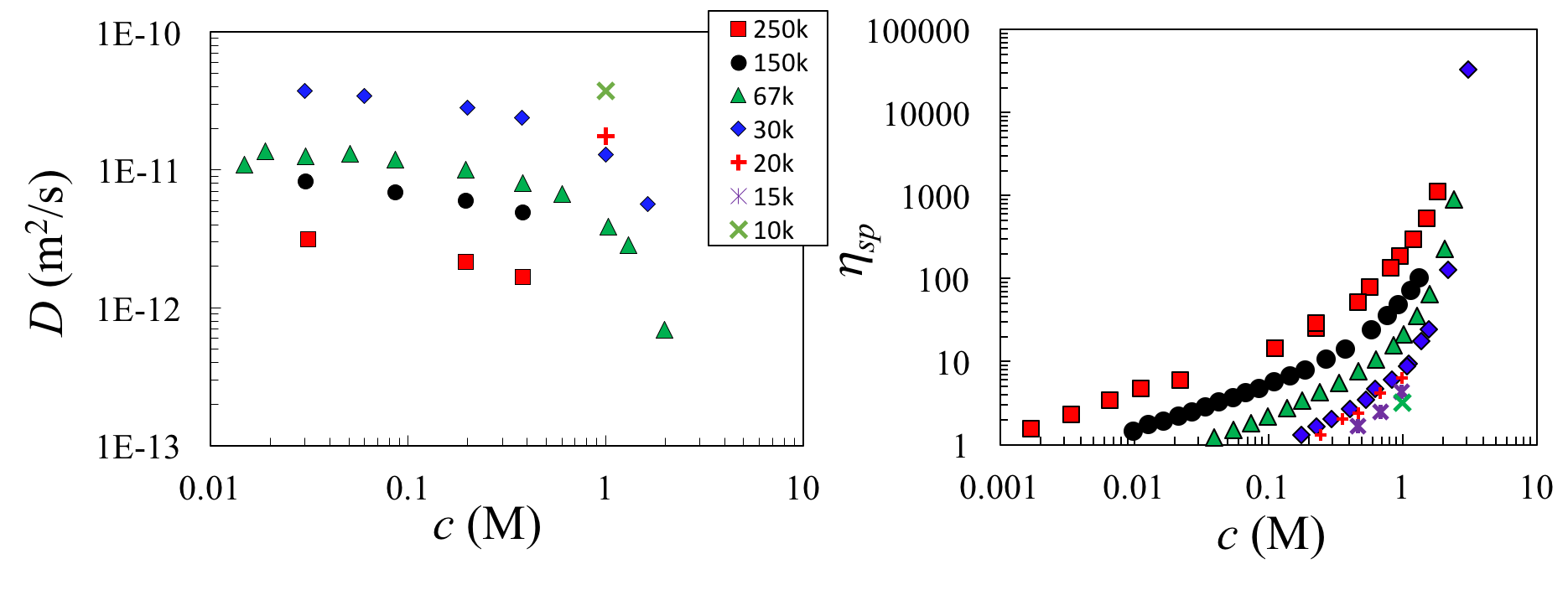}\vspace{-0.7cm}
\caption{Left: Diffusion coefficient of  NaPSS samples in salt-free D$_2$O  as a function of polymer concentration. Right: Specific viscosity for the same NaPSS samples in H$_2$O. Viscosity data for samples with $M_w = 29.4-261$kg/mol are from ref. [\citenum{PSSII}]. }
\label{fig:D}
\end{figure*}

PFG-NMR can be used to study  molecular diffusion in solution\cite{guo2017diffusion,williamson2017scaling}. The PFG-NMR technique combines the ability to reveal information on the chemical nature as well as on the molecular or collective translational mobility of the individual components. The observed molecules, may be assigned to a structural part of the dispersion depending on its characteristic motional behaviour. \cite{NMR1,NMR2} In a homogeneous solution, the free self-diffusion coefficients of the dissolved molecules can be determined with the PFG-NMR. In the given case, all PFG-NMR experiments consist of the application of two field gradient pulses combined with a stimulated echo pulse sequence (90$^\circ$-$\tau$1-90$^\circ$-$\tau$2-90$^\circ$-$\tau$1-echo). The pulse gradients with a gradient strength $G$ and duration $\gamma$ are applied during both of the waiting periods $\tau$1 with an overall separation $\Delta$. In the presence of free diffusion with a diffusion coefficient $D$, this leads to a decay of the echo intensity $I$ with respect to the original value $I_0$ (for $G = 0$) according to:\vspace{-0.2cm}
$$
\frac{I}{I_0} = I_{rel} = e^{-\gamma^2\delta^2G^2(\Delta -\delta/3)}
$$
\noindent with $\gamma$ being the gyromagnetic ratio of the hydrogen nucleus. The  value of the apparent diffusion constant $D_{app}$   is equal to the negative slope of the data points  in the Stejskal-Tanner plot (ln $I_{rel}$ versus $\gamma^2\delta^2G^2(\Delta -\delta/3))$.\cite{NMR1,NMR3,NMR4}

Plots of  the decay of the echo intensity for the NaPSS sample with $M_w = 63.9$ kg/mol for different polymer concentrations in salt-free D$_2$O are shown in Figure \ref{fig:D}. The DOSY spectra for the sample is shown in the SI. %

\textcolor{black}{The viscosity data did not display any appreciable shear thinning over the shear rate range probed (1-600s$^{-1}$), as expected given the relatively short relaxation times of the low molar mass polymers studied\cite{PSSI}.}

\section{Discussion}\label{sec:discussion}\vspace{-0.4cm}

\subsection{Calculation of $\psi(c)$}\vspace{-0.3cm}
Figure \ref{fig:D} displays the diffusion coefficient  and specific viscosity of NaPSS  as a function of polymer concentration in salt-free aqueous  solution. The data follow the scaling predictions at low concentrations $D \propto c^0$ and $\eta_{sp} \propto c^{1/2}$, followed by a sharp  decrease in $D$ and increase in $\eta_{sp}$ in the high concentration region. 
\begin{figure}[h!]
\includegraphics[width=3.3in]{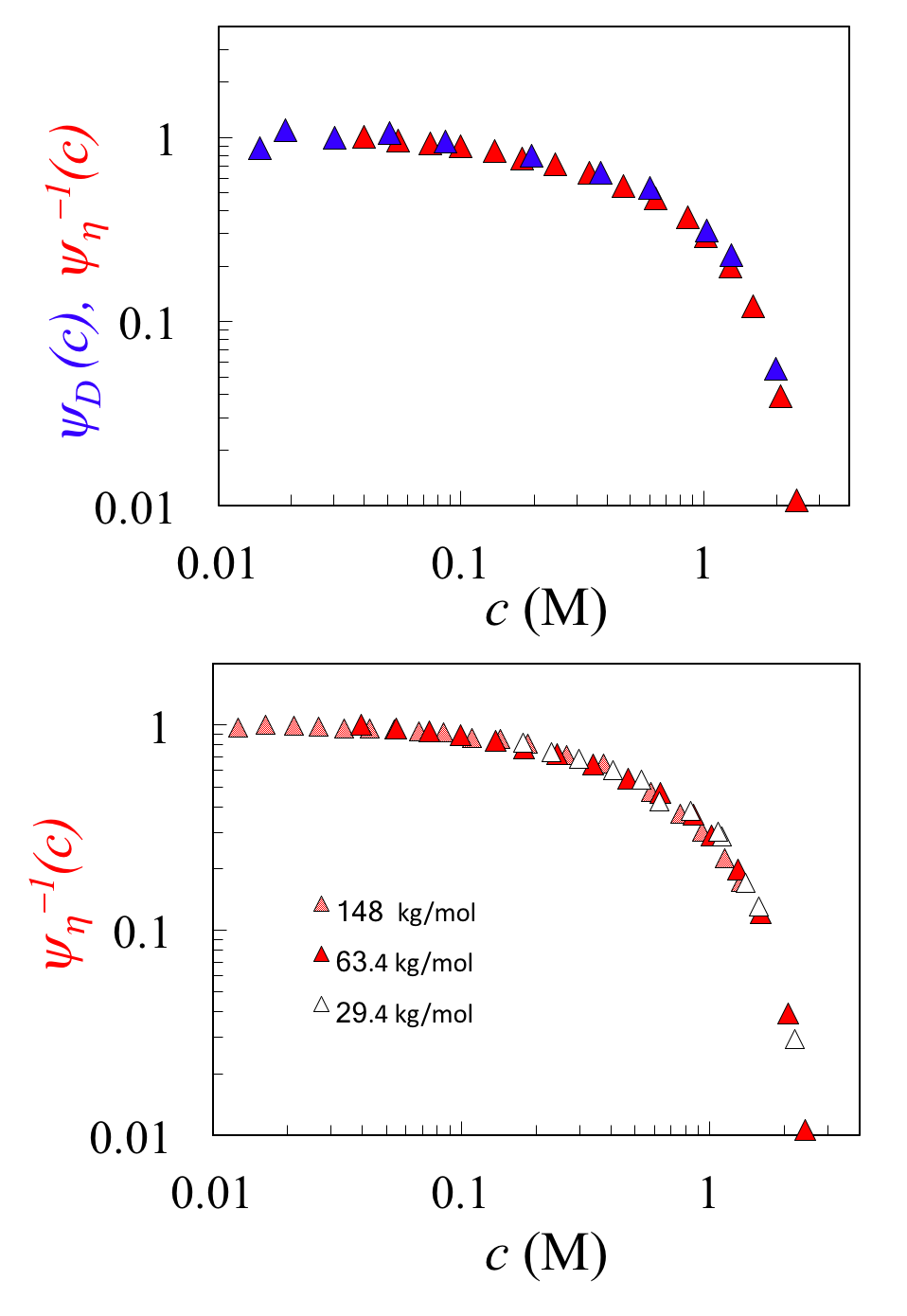}\vspace{-0.7cm}
\caption{Top: Comparison of $\psi_\eta$ and $\psi_D$ for PSS with $M_w = 6.34 \times 10^4$ g/mol. Bottom: Concentration dependence of $[\psi_\eta(c)]^{-1}$ for PSS with different molar masses.}\vspace{-0.3cm}
\label{fig:Psi1}
\end{figure}

As discussed in the introduction, the simplest explanation for the observed deviations from scaling, which are quantified by the function $\psi(c)$ is to assume a concentration-dependent friction coefficient $\beta$. This interpretation leads to two requirements: First, following Eqs. \ref{eq:DRouse}, \ref{eq:etaRouse} and \ref{eq:psi}, a change in local friction is expected to affect the diffusion coefficient and viscosity inversely, so that $\psi_D(c) = \psi_\eta(c)^{-1} = \beta(c)^{-1}$. Second, because local properties such as the monomeric friction coefficient are insensitive to the overall chain length, the shape of $\psi(c)$ must be $N-$independent. Note that the second requirement does not necessitate that $\psi(N) = 1$, but  $\psi(c)/\psi(c\to0) \propto N^0$. An alternative interpretation, were $F$ is assumed to vary with concentration, is discussed in the next section.

Figure \ref{fig:Psi1}a compares $\psi_D(c)$ and $\psi^{-1}_\eta(c)$ for NaPSS samples with $M_w = 6.39 \times 10^4$ g/mol. The agreement between the two functions is superb. This contrasts with the findings of an earlier study\cite{PSSII}, where the same viscosity data were compared with diffusion data by Odijk and Ooswald\cite{oostwal1993novel}. 

The independence of $\psi(c)/\psi(0)$ on $N$ is verified in Fig. \ref{fig:Psi1}b for $\psi_\eta(c)$ of three different molar masses. $\psi_{\eta}(c)$ is calculated by dividing experimental data for $\eta_{sp}$ by Eq. \ref{eq:etaRouse} and adjusting the lowest concentration to $1$ for the $M_w = 1.48 \times 10^5$ g/mol sample. For the $M_w = 2.94 \times 10^4$ and $6.39 \times 10^4$ g/mol samples, data at sufficiently low concentrations to observe a plateau in $\psi_\eta(c)$ are not available and we therefore normalise the lowest concentrations to the  $M_w = 1.48 \times 10^4$ g/mol sample. Very good agreement is observed in Fig. \ref{fig:Psi1}b over the entire concentration range considered, supporting the $\psi_\eta(c) = \beta(c)$ interpretation.

\subsection{The shape factor $F$}
\subsubsection{Hydrodynamic models}

A quantitative comparison of scaling laws with dynamic data for polyelectrolytes requires a precise knowledge of $F$. As the quantities $\beta$ and $F$ only appear as a product in the scaling laws outlined in the introduction, their individual evaluation is challenging. We proceed as follows: at low polymer concentrations, we suppose that solvent dynamics are not modified by the polymer, which in turn means that $\beta = \eta_s$ is a reasonable assumption. Using previously reported SANS measurements of the correlation length\cite{prabhu2003polyelectrolyte} and effective monomer size\cite{VdM}, we estimate the friction coefficient of a correlation blob as $\zeta_\xi = \zeta_R\xi/(b'N) = \zeta_R (33c^{-1/2})/(1.7N)$.\cite{PSSI} Figure \ref{fig:F} plots the concentration dependence of  $\zeta_\xi$ for NaPSS samples with $M_w =$ 148 and 29.1 kg/mol. Our data for other molar masses and Oostwal and co-workers's data\cite{oostwal1993novel,oostwal1993chain} for  fall in between these values, and are generally closer to the triangles, see the supporting information for more details. A comparison is made with the friction coefficients calculated Eq. \ref{eq:Fchain} using the scaling value of $F = 1$ and that of a cylinder, which is given by\cite{tirado1979translational}:

\begin{figure}[h!]
\includegraphics[width=3.3in]{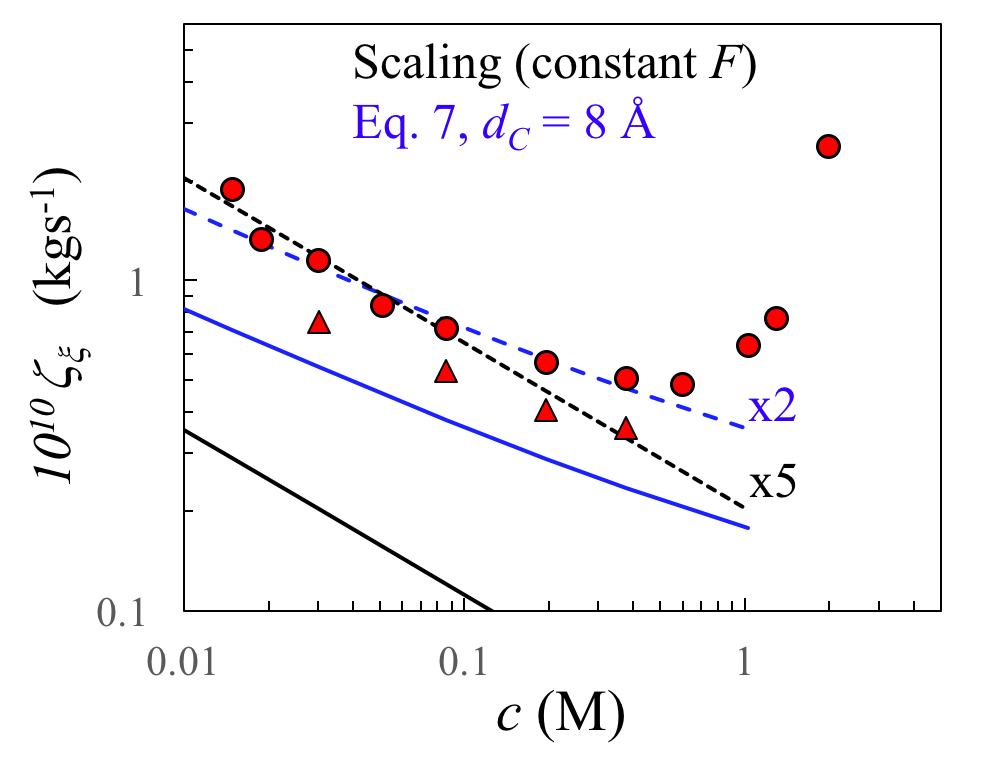}
\caption{Comparison of friction coefficient of correlation blobs Eq. \ref{eq:Fchain} for different values of $F$. Full blue lines is for $F$ calculated with Eq. \ref{eq:Fcyl} ($L = \xi$, $d_C = 8$\AA) and full black line is for $F = 1$, as expected by scaling. Dashed lines are multiplied by factors indicated to match data for NaPSS with $M_w = 29.4$ kg/mol.}
\label{fig:F}
\end{figure}

\begin{equation}
\begin{split}
F_{cyl}(L,d_C) \simeq 3\pi \Big[ln\Big(\frac{L}{d_C}\Big) + 0.312 + 0.565\Big(\frac{d_C}{L}\Big) \\  -0.1\Big(\frac{d_C}{L}\Big)^2\Big]^{-1}
\end{split}
\label{eq:Fcyl}
\end{equation}

\noindent where $L$ and $d_C$ are the length and the cross-sectional diameter of the cylinder respectively. The former can be identified with the correlation blob and the latter is taken to be 10 \AA.\cite{VdM}  

Using either the scaling value of $F = 1$ or $F_{cyl}(\xi,d_C)$ fail to describe the data quantitatively. Setting $F = 4-5$ instead correctly describes experimental results for $c \simeq 0.02-0.2$, see also Table \ref{tab:F}.  At higher polymer concentrations, the downturn in the friction coefficient can be assigned as discussed in the preceding section, to an increase in $\beta$ with increasing polymer concentration.

Alternatively, using Eq. \ref{eq:Fcyl} could account for results if $F_{cyl}$ is multiplied by a factor of 1.4-2 (see dashed line). Under this scenario, $\psi(c)_D$ for $c \lesssim 0.4$ would be entirely explained by the concentration dependence of $F$, with $\beta \simeq \eta_s$ in that concentration range. Inserting Eq. \ref{eq:Fcyl} into Eq. \ref{eq:etaRouse} predicts a power-law exponent for the specific viscosity with concentration stronger than 0.5 at low polymer concentrations, which is inconsistent with the experimental data for the viscosity of NaPSS \cite{ColbyPSS,PSSI,PSSII,PSSIII}. Conductivity data\cite{ColbyConductivity,DielectricReview,de2007effects,ray2016thermodynamic} on the other hand display a logarithmic dependence on polymer concentration in dilute solution, which is  interpreted as arising from the first term in the square brackets of Eq. \ref{eq:Fcyl}. 

It is possible to reproduce the experimental value of $F \simeq 5$  with the cylinder friction factor if a constant $\xi/d_C \simeq 10$ is assumed. Such behaviour could arise if, for example, the ionic cloud around the polyelectrolyte backbone causes the hydrodynamic cross-sectional radius of the chain to increase with decreasing concentration as $d_C \simeq 1/\sqrt{c}$.

\begin{table}[h!]
\begin{tabular}{l|p{5.2cm}|l}
Model & $F$  & $F_{exp}/F$\\ \hline
Sphere & $6\pi$ & $\simeq$4\\
Cylinder & $3\pi[ln(\xi/d_C) + 0.312 + 0.565(d_C/\xi)  -0.1(d_C/\xi)^2]^{-1}$ & 1.7-2.4\\
Zimm & 5.2$^a$ & 1\\
Scaling & 1 & 5\\
Experiment & $\simeq 5 \pm 1$ & 1
\end{tabular}
\caption{Comparison of theoretical and measured values for shape factor $F$, see figure \ref{fig:F} for the experimental determination of $F$. $^a$ for Gaussian chains with  $R = \xi$.}\vspace{-0.4cm}
\label{tab:F}
\end{table}

\subsubsection{Comparison of dilute and semidilute chain diffusion}\vspace{-0.4cm}

\begin{figure}[h]
\includegraphics[width=3.2in]{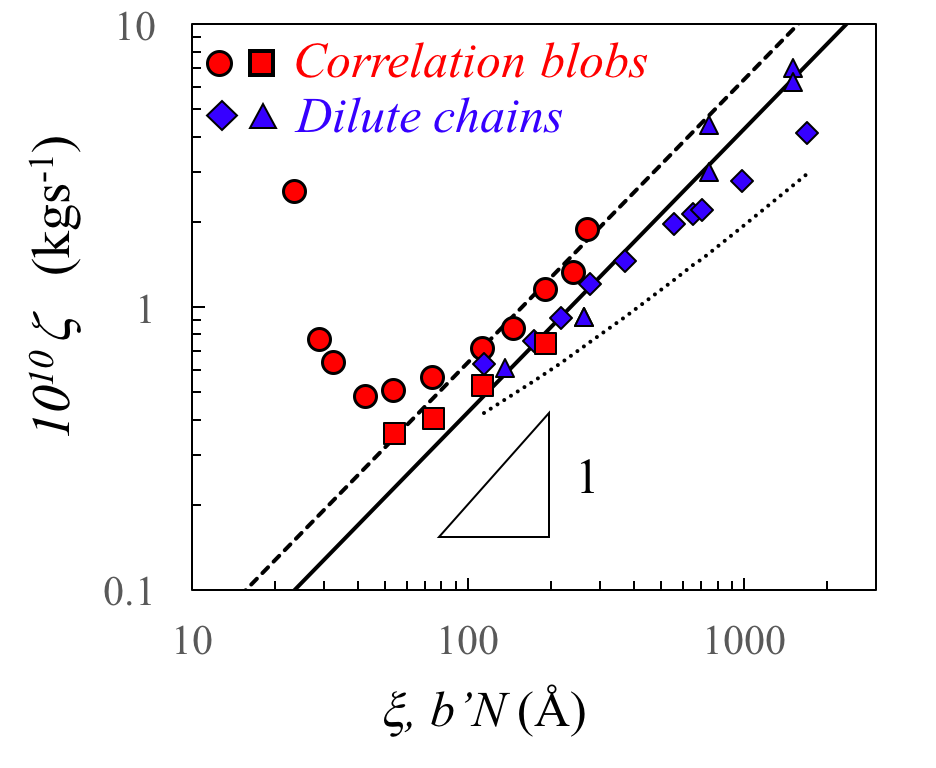}
\caption{Comparison of friction coefficient of correlation blobs with that of dilute chains of the same end-to-end distance. Blue symbols are for dilute solutions of NaPSS in salt-free water: Triangles are Oostwal and co-worker's data, extrapolated to zero concentration and diamonds are  Xu et al's fluorescence measurements\cite{xu2016single,xu2018molecular,chen2019diffusive}. Red symbols are $\zeta_\xi$ for NaPSS with $M_w = 29.1$ kg/mol (circles) and $M_w = 248$ kg/mol (squares), full lines are power-laws of 1, predicted by scaling and dotted line is the diffusion coefficient of a cylinder with $d_C = 10 $\AA.}
\label{fig:zetaxi}
\end{figure}

The scaling assumption to calculate polymer dynamics is that chains display dilute-like behaviour (Zimm dynamics) for distances smaller than the correlation length and melt-like behaviour (Rouse dynamics) on larger length-scales. Figure \ref{fig:zetaxi} tests this assumption by comparing the friction coefficient of correlation blobs (red symbols) with those of dilute chains of the same length (blue symbols). Both datasets are seen to differ by a factor of $\simeq 1-1.5$, supporting the validity of the scaling theory.

\subsection{Calculation of $\psi(N)$}
\begin{figure}[h!]
\includegraphics[width=3.3in]{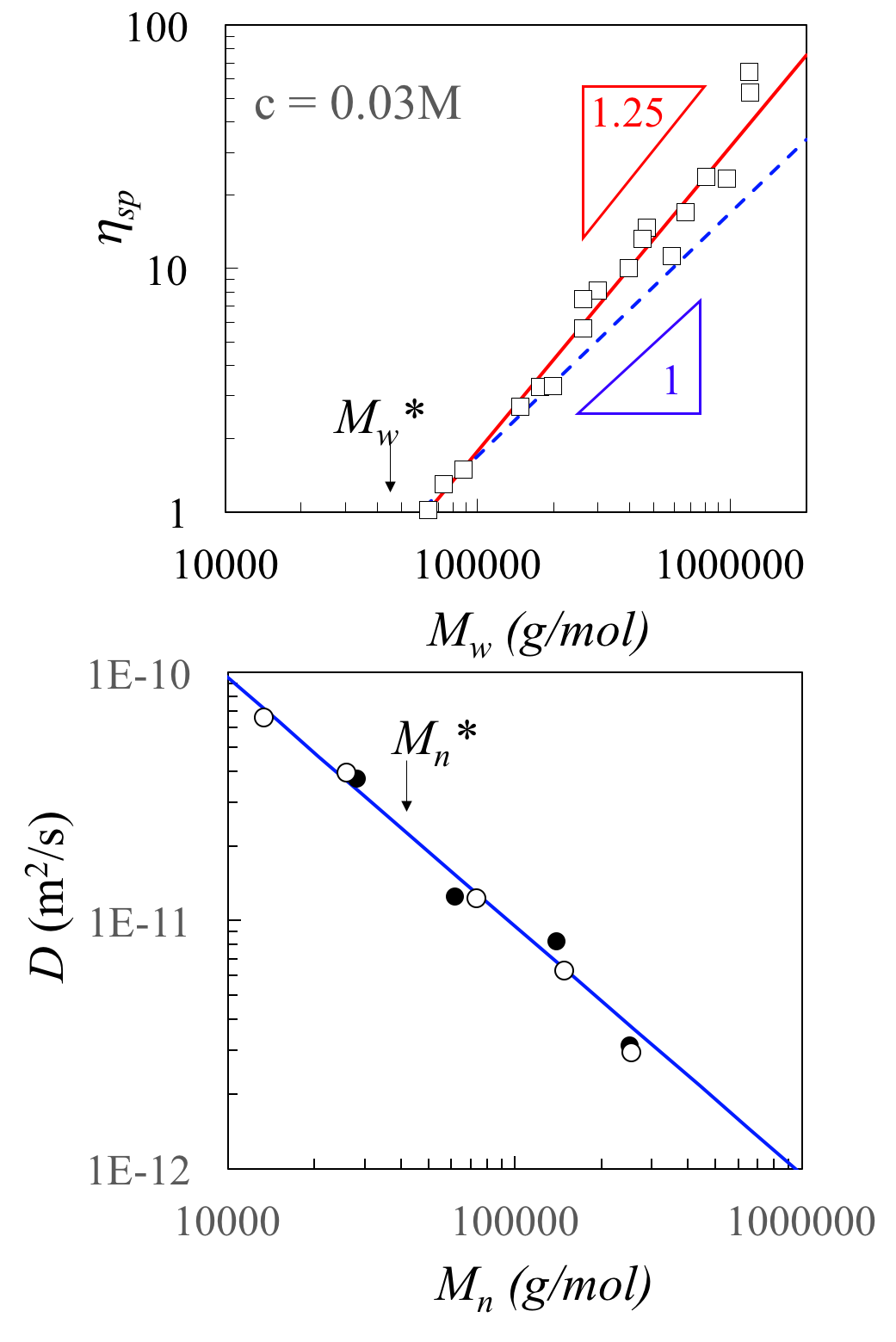}
\caption{Top: $M_w$ dependence of specific viscosity for NaPSS in salt-free water with $c = 0.03$ M, data are from this work and literature references compiled in ref \citenum{PSSI}. Bottom: Diffusion coefficient as a function of number-averaged molar mass for the same concentration as top panel. Full symbols are data from this work and hollow symbols are by Oostwal et al\cite{oostwal1993novel}.}
\label{fig:etaD1}
\end{figure}

\vspace{-0.5cm}
Figure \ref{fig:etaD1}a compares the dependence of the specific viscosity on the polymer molar mass with the Rouse prediction of $\eta_{sp} \propto M_w$. As discussed in earlier work\cite{PSSI}, a power-law of $\eta_{sp} \propto M_w^{1.25}$ is seen to give a better description of the experimental results over the entire $M_w$ range. An earlier study has shown that this power-law persists well into the dilute region, which is at odds with the scaling assumption of Zimm dynamics in dilute salt-free solution.\cite{PSSIII} The diffusion coefficient, plotted as a function of $M_n$ in Fig. \ref{fig:etaD1}b, appears to follow the Rouse prediction of $D \propto M^{-1}$, with no apparent change in scaling across the overlap degree of polymerisation. The $D$ vs. $M_n$ dataset does not cover a sufficiently broad range of degree of polymerisation to distinguish between a power-law of $D \sim N^{-1}$ or $D \sim N^{-1.2}$. The same dependences are observed at higher concentrations of up to $c = 1$ M, see the SI. 

\subsection{Comparison with models of solvent dynamics}\vspace{-0.4cm}

The preceding discussion suggests that the sharp drop in $\psi_D(c) \simeq \psi^{-1}_\eta(c)$ cannot be accounted for by either the influence of entanglements or hydrodynamic corrections to the shape factor $F$. 

The  simplest explanation therefore appears to be a slow down in the  solvent diffusion with increasing polymer concentration. This phenomenon  has been observed for many  polymer solvent systems, and might be particularly strong for polyelectrolytes due to the ordering of water around the ionic groups \cite{quezada2020molecular,sappidi2014conformations,kubincova2020interfacial,schlaich2019simulations}. In this scenario the scaling laws outlined in section \ref{sec:Scaling} need to be applied with $\beta(c) > \eta_s$. Several theories of solvent diffusion in polymer solutions have been put forward in the literature. Mackie and Meares\cite{MM} developed an obstruction model which expects the solvent diffusion to depend on on the polymer volume fraction ($\phi$) as:
 \begin{equation}
\frac{D_s(\phi)}{D_s(0)} = \Big[\frac{1-\phi}{1+\phi}\Big]^{2}
\label{eq:MM}
\end{equation}

Equation \ref{eq:MM} has been shown to correctly describe the dependence of $D_s$ on polymer concentration for a large number of polymer-solvent pairs, including polyelectrolytes\cite{bai2014nmr}.

Fujita introduced a  model where the free volume available per molecule is estimated as an additive contribution from solvent and solute\cite{Fujita1991}:

\begin{equation}
f_v(\phi,T) = \phi f(1,T) + (1-\phi)f(0,T)
\label{eq:FV}
\end{equation}

\noindent where $f_v$ is the total free volume of the solution and $f(1,T)$ and $f(0,T)$ are  the free volume of the pure polymer and solvent respectively. Models based on Fujita's concept of additive free volume predict: 
\begin{equation}
D \sim e^{-B/f_v}
\label{eq:DFV}
\end{equation}
\noindent where $B$ is often taken to be a constant of order unity. The Vrentas-Duda model gives a similar prediction, but $B$ is a function of polymer concentration $B =  (1-\phi)/\rho_s + k_1\phi/\rho_p$, where $\rho$ is the density and the subscripts $p$ and $s$ refer to the polymer and solvent. $k_1$ is a constant specific to each polymer-solvent system.\cite{sturm2019application}

\begin{figure}[h!]
\includegraphics[width=3.4in]{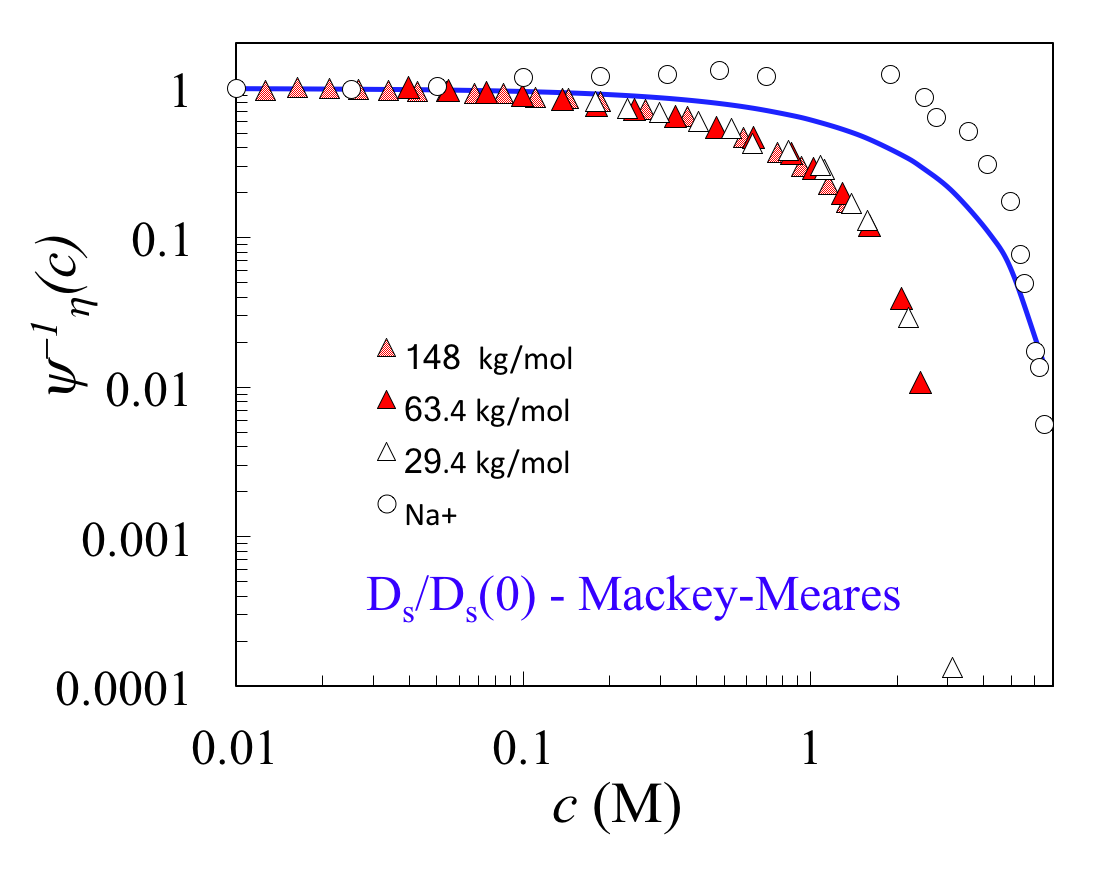}
\caption{Concentration dependence of $\psi_\eta^{-1}$ for NaPSS with different molar masses along with Eq. \ref{eq:MM} (blue line). The diffusion coefficient of Na$^+$ counterions normalised to its value at $c = 0$ is plotted as circles, data are from refs. [\citenum{prini1964tracer,fernandez1976tracer}].}
\label{fig:FV1}
\end{figure}

Figure \ref{fig:FV1} compares the concentration dependence of $\psi_{\eta}$ with Eq. \ref{eq:MM}. The latter is shown to predict a much weaker dependence of solvent dynamics on concentration than what is observed for NaPSS chains. Free volume models give a similarly weak dependence of the friction coefficient with concentration. The various terms in Eq. \ref{eq:FV}-\ref{eq:DFV} could in principle be evaluated from the temperature dependence of the viscosity as a function of polymer concentration. Such data have proven useful when interpreting the dynamics of salts, proteins, nanoparticles and polymers in solution.\cite{DouglasNP,DouglasHF,DouglasProtein,colby1991effects} Unfortunately, with our current setup, evaporation prevents us from measuring the solution viscosity over a wide enough temperature range to evaluate our data according to the procedures outlined in refs \cite{gong2001self,schneider2010steady,colby1991effects}.

The diffusion coefficient of the sodium counter-ion in salt-free solutions of NaPSS, reported in refs. [\citenum{fernandez1976tracer,prini1964tracer}] does not show a concentration dependence up to $c \simeq 2$ M, beyond which a sharp drop, similar to that of NaPSS chains is observed. The independence of $D_{Na^+}$ on concentration up to $c \simeq 2$ M supports the idea of a weak concentration dependence of the solvent diffusion.

Other effects, including the influence of coupled chain motion \cite{chen2019diffusive} or hydrophobic clustering\cite{zhang2016study} may affect the shape of the $\psi(c)$ function, but we do not have a way of quantifying them at present. At high polymer mass fractions, the dielectric constant of polyelectrolyte solutions varies non-monotonically with concentration\cite{DielectricReview,bordi2002electrical}, which is expected to affect polyelectrolyte conformation\cite{liu2017shape}. This feature, along with a possible breakdown of Debye-Huckel  screening\cite{jacinto2018static,matsumoto2019rheological} may be related to the observed increase in the osmotic coefficient for NaPSS and other polyelectrolytes beyond $c \simeq 1.2$ M,\cite{BloomfieldOsmotic,ullner2018osmotic} concurrent with a change in the scaling of the correlation length from $\xi \sim c^{-1/2}$ to $\xi \sim c^{-1/4}$.\cite{NishidaHighC,Lorchat2014} While these  observations  suggest a cross-over to a concentrated regime, the conformation of NaPSS displays a  monotonic decrease in $R_g$ with increasing polymer concentration up to $c \simeq 4$ M, which closely the predictions of the scaling theory\cite{PSSI}.

In summary, the above observations suggest that the monomeric friction factor of NaPSS exhibits a strong concentration dependence beyond $c \simeq 0.1$ which cannot be assigned to a change in solvent dynamics or the influence of entanglements. Interchain friction  therefore appears as the most likely cause for the strong exponents observed for the viscosity and diffusion coefficient of NaPSS.

Our findings are consistent with  data  for sodium polymethacrylate (NaPMA) in non-entangled, salt-free D$_2$O solution\cite{schipper1999mass,NotePMA}, which show a decrease in the single chain diffusion coefficient of of NaPMA relative to that of the solvent for $c \gtrsim 0.05$ M. The $\psi_\eta(c)$ function calculated for their data shows  agreement with our results, see the supporting information. Diederichsen et al's data for the diffusion  of polysulfone ionomers (PSU) in DMSO\cite{diederichsen2019counterion,diederichsen2019high} display qualitatively similar behaviour to the data presented in this paper: the viscosity and diffusion coefficient of PSI display a sharp increase and decrease respectively at high polymer concentrations. These features cannot be fully accounted for by  changes in solvent dynamics, as their measurements show that DMSO diffusion slows down only modestly  at high polymer concentrations. Entanglement can also be ruled out given the low degree of polymerisation of the PSU chains and the observed exponents of $D \sim N^{-(1.1-1.2)}$ and $\eta_{sp} \sim N^{1-1.2}$.

\subsection{Estimating chain dimensions from viscosity and diffusion data}

The Rouse model relates the viscosity, diffusion and chain dimensions of non-entangled chains as:

\begin{equation}
k_BT  \frac{R_g^2}{6N} = \frac{(\eta-\eta_s)D}{c}K
\label{eq:RS}
\end{equation}

\noindent where  $K$ is a factor that accounts for polydispersity effects\footnote{If $D$ is measured from the initial slope of the PFG-NMR signal (i.e. $D = \lim_{q^2 \to 0} d[S(q)/S(0)]/dq^2$), then $K \simeq M_w/M_n$ for semidilute non-entangled solutions}. The scaling theory gives the same result as Eq. \ref{eq:RS} but with the left hand side multiplied by a factor of 36. Equation \ref{eq:RS} has been shown to work within 10-20\% accuracy for non-entangled neutral polymers in concentrated solutions and in the melt. \cite{nemoto1991self,nemoto1989self}

\begin{figure}[h!]
\includegraphics[width=3.4in]{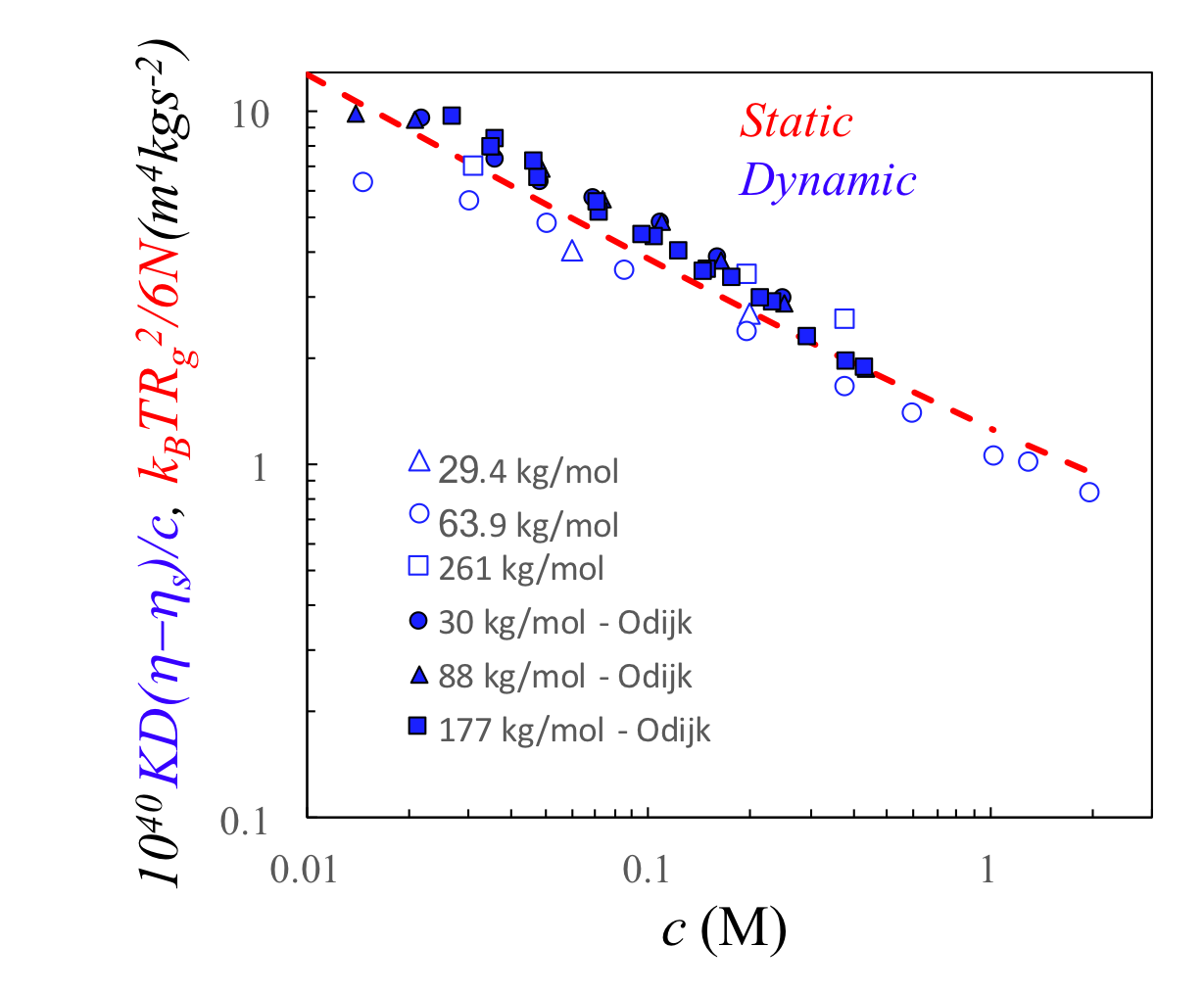}
\caption{Concentration dependence of the chain dimensions of NaPSS in salt-free solution calculated from Eq. \ref{eq:RS}. SANS data follow the compilation of \citenum{PSSI}, data for $c <$ 0.07 M are extrapolated.}
\label{fig:OO1}
\end{figure}

The advantage of Eq. \ref{eq:RS} is that it relates static ($R_g$) and dynamic ($D$, $\eta$) experimental observables in a way in which the effects of local friction, quantified through parameter $F\beta$, are cancelled out. \footnote{we have multiplied all diffusion coefficients by a factor of 1.2 to account for the fact that the viscosity of D$_2$O is $\simeq$ 20\% larger than that of H$_2$O at $T = 298$ K.} Figure \ref{fig:OO1} compares the chain dimensions of NaPSS calculated from Eq. 9 (blue symbols) with those directly measured by SANS, see ref. \citenum{PSSI} for a compilation of the SANS data. Very good agreement is observed between the two methods over the entire concentration range studied, demonstrating the validity of the Rouse model for salt-free polyelectrolytes. Table \ref{tab:R2} compares  estimates for the chain dimensions of NaPSS using the Rouse model, the scaling formula based on the correlation length and direct measurements by SANS. The three estimates are seen to agree within experimental error.

\begin{table}[h!]
\begin{tabular}{l|p{3cm}l}
Method &$R_g^2c^{0.5}/N$ (\AA$^2$)   \\ \hline
$\xi(L/\xi)^2/6$$^\dag$ & $9 \pm 1$\\
$(\eta-\eta_s)DK/(6ck_BT)$ & 14 $\pm$ 3 \\
SANS & 12 $\pm$ 2
\end{tabular}
\caption{Comparison of calculated and measured values for chain dimensions of NaPSS. $^\dag$ Assuming and effective monomer length of $b' = 1.7$ \AA. For a discussion of SANS data see ref. \cite{PSSI} and references therein.}
\label{tab:R2}
\end{table}

\section{Conclusions}
We have studied the diffusion and viscosity properties of NaPSS in salt-free aqueous solution as a function of degree of polymerisation and concentration. We show that while $D$ and $\eta$ depart from scaling predictions, their product does not, suggesting that the observed differences between theory and experiments are due to a concentration dependence of the friction coefficient of correlation blobs. Free volume theories\cite{Fujita1991} and obstruction models\cite{MM} do not account for this dependence. The friction coefficient of correlation blobs and dilute chains of equivalent length are shown to be nearly identical, in agreement with scaling.\cite{Dobrynin} The Rouse model gives a quantitative description of non-entangled polyelectrolyte dynamics, as evidenced by the fact that $(\eta - \eta_s)D//c$ can be used to calculate chain dimensions in agreement with SANS measurements. 

Overall, scaling gives a good description of non-entangled polyelectrolyte dynamics. One exception is the exponent for the molar mass dependence of the specific viscosity in salt-free solution, which deviates from theoretical predictions.\cite{PSSI} Further, a molecular understanding of the concentration dependence of the friction coefficient is also still lacking. 

\section*{Acknowledgements}
We thank Jack Douglas (NIST, US) for useful comments on the manuscript.

\section{References}
\nocite{apsrev41Control}
\bibliography{PSSDiffusion}

\begin{thebibliography}{145}
\bibitem [{Not({\natexlab{b}})}]{NotePMA}%
  \BibitemOpen
  \href@noop {} {\enquote {\bibinfo {title} {We assume their {NaPMA} is non
  -entangled given its low degree of polymerisation ${N} = 90$.}}\ }
  ({\natexlab{b}})\BibitemShut {NoStop}%
\end{thebibliography}

\end{document}